\documentclass[12pt]{article}

\usepackage{mathptmx}
\usepackage{graphicx}

\usepackage{amsfonts}

\begin{document}

\title{Superextension of Jordanian Deformation \\
for $ U(osp(1|2))$ \\  and
its Generalizations }

\author{A. Borowiec$^\dag$ , J. Lukierski$^\dag$ \  and V.N.
Tolstoy$^\ddag$
\\ \\
$^\dag$Institute for Theoretical Physics, \\
 University of Wroc{\l}aw,
pl. Maxa Borna 9,                             \\
 50--205 Wroc{\l}aw, Poland
\\      \\
$^\ddag$Institute of Nuclear Physics,
  \\
Moscow State University, 119992 Moscow, Russia}

\date{}

\maketitle

\begin{abstract}
We describe Jordanian ``nonstandard'' deformation of $U(osp(1|2))$ by employing
 the twist quantization technique. An extension of these results to $U(osp(1|4))$
 describing deformed
  graded $D=4$  $AdS$ symmetries and to  their super-Poincar\'{e} limit is outlined.
\end{abstract}

\section{Introduction}
It is well-known that the $U(sl(2))$ algebra with  basic
commutators

\begin{equation}\label{lume1}
[h,e_{\pm}] = \pm e_{\pm} \, , \qquad [e_+, e_- ] = 2h \, ,
\end{equation}
can be endowed with the following two inequivalent quantum
deformations:

i) Drinfeld-Jimbo ``standard'' $q$-deformation with the following
classical $r$-matrix ($q=1- \gamma $)
\begin{equation}\label{lume2}
    r_{_{\!DJ}} =\gamma\,e_{+}^{}\wedge e_{-}^{}\, ,
\end{equation}

ii) Jordanian ``nonstandard'' quantum deformation, generated by
the classical $r$-matrix
\begin{equation}\label{lume3}
    r_{\!_{J}}^{}=\xi\,h\wedge e_{+}^{}\,.
\end{equation}
The Hopf-algebraic structure of the Drinfeld-Jimbo deformation $U_q(sl(2))$ was
given firstly in \cite{bor1}--\cite{bor3},
 and the Hopf algebra describing Jordanian
deformation of $U(sl(2))$ was presented in  \cite{bor4}.

The classical $r$-matrix (\ref{lume3}) satisfies classical YB
equation and its quantization can be described by so-called twist
quantization method \cite{bor6}.
 We recall that twist quantization of a
Hopf algebra $H=(A, m, \Delta, S, \epsilon)$ is given by the twisting
two-tensor $F=\sum\limits_{i} f^{(1)}_{i} \otimes   f^{(2)}_{i}$
modifying the coproduct $\Delta$ and antipode $S$ as follows:

\setcounter{equation}{0}
\renewcommand{\theequation}{4\alph{equation}}

\begin{eqnarray}\label{lume4a}
&&  \Delta \to  \Delta_F
=F\, \Delta\, F^{-1}\, ,
\\ \label{lume4b}
&& S \to S_F=u\,S\, u^{-1},
\quad\;u= \sum_i f^{(1)}_{i}\, S\, (f^{(2)}_i)\, .
\end{eqnarray}
It should be stressed that the algebraic sector $A$ of $H$ remains unchanged.

The twisting two-tensor $F$ for the Jordanian deformation of
$U(sl(2))$ (see (\ref{lume3})) has been given firstly by Ogievetsky
 \cite{bor7}
  in the following closed form

\setcounter{equation}{4}
\renewcommand{\theequation}{\arabic{equation}}

\begin{equation}\label{lume5}
    F_{\!_{J}}=\exp(\xi\,h \otimes E_{+}^{})\, ,
\end{equation}
where
\begin{equation}\label{lume6}
    E_{+}^{}=\frac{1}{\xi}\,\ln(1+\xi\,e_{+}^{})=e_{+}^{}+{\mathcal O}(\xi)\, .
\end{equation}

The deformations (\ref{lume2}) and (\ref{lume3}) of $U(sl(2))$ provide important building
blocks in the theory of quantum deformations of arbitrary Lie algebra. Similar
role in the deformation theory of Lie superalgebras is played by the deformations of
rank 1 superalgebra $osp(1|2)$, which is the  supersymmetric extension of
$sl(2) \simeq sp(2)$. Our first aim here is to generalize the Jordanian deformation of
$U(sl(2))$ to the $U(osp(1|2))$ case. Further we present briefly the Jordanian deformation
of $U(osp(1|4))$ as a special example of general framework presented by one of
the authors in \cite{bor8}.
 By interpreting $osp(1|4)$ as  $D~=~4$ AdS superalgebra we were able to obtain
via contraction a new $\kappa$-deformation of  $D~=~4$ Poincar\'{e} superalgebra  \cite{bor9}.

\section{Jordanian Deformation of $U(osp(1|2))$}

The classical $r$-matrices (\ref{lume2}) and (\ref{lume3}) are
supersymmetrically extended as follows:
\setcounter{equation}{0}
\renewcommand{\theequation}{7\alph{equation}}

\begin{eqnarray}\label{lume7a}
    r_{_{DJ}}^{susy}&=&\gamma (e_{+}^{}\wedge e_{-}^{}
    + 2 {v}_{+}^{}\wedge v_{-}^{})\,,
\\[9pt] \label{lume7b}
r_{_{J}}^{susy}&=& \xi(h\wedge e_{+}^{}-v_{+}^{}\wedge v_{+}^{})\,,
\end{eqnarray}
where the odd generators $v_{\pm}$ of $osp(1,2)$ extend the
$Sl(2)$  algebra (\ref{lume1}) as follows:
\setcounter{equation}{7}
\renewcommand{\theequation}{\arabic{equation}}

\begin{eqnarray}\label{lume8}
[h,\,v_{\pm}^{}]=\pm\,\frac{1}{2}\,v_{\pm}\,, \quad\!\!\!&&\!\!\!\quad
\{v_{+}^{},\,v_{-}^{}\}=-\frac{1}{2}\,h\,,
\nonumber \\[7pt]
 e_{\pm}^{}\!\!&=&\!\!\pm4\,(v_{\pm}^{})^2\, ,
\end{eqnarray}
and in (\ref{lume7a}--\ref{lume7b})
  for odd generators we define $a \wedge b = a \otimes b + b \otimes a$.

The quantization of the deformation (\ref{lume7a}) is well known \cite{bor10}
 as a particular case of the extension of Drinfeld-Jimbo quantization method to
 Lie superalgebras \cite{bor11,bor11a}.
  The Jordanian quantization of $U(osp(1|2))$ generated
 by (\ref{lume7b}) has been obtained quite recently, by the superextension of twist
 quantization procedure \cite{bor12}.
  It should be mentioned that incomplete discussion
 of twist quantization of $U(osp(1|2))$  was presented earlier
 \cite{bor13,bor14},
  but explicite formulae for the
  twist tensor and all coproduct formulae have been given firstly in \cite{bor12}.

 Let us recall that twisting element $F$ should satisfy the cocycle
 equation
 \begin{equation}\label{lume9}
    F^{12}(\Delta\otimes{\rm id})(F)=F^{23}({\rm id}\otimes\Delta)(F)\,,
\end{equation}
and the ``unital" normalization condition
\begin{equation}\label{lume10}
    (\epsilon \otimes{\rm id})(F)=({\rm id}\otimes\epsilon )(F)=1\,.
\end{equation}
We assume that the twisting two-tensor $F_{_{\!SJ}}$ describing the quantization
of the classical $r$-matrix (\ref{lume7b}) can be factorized as follows
\begin{equation}\label{lume11}
    F_{_{\!SJ}} = F_{_{\!S}}\, F_{_{\!J}} \, ,
\end{equation}
where the ``supersymmetric part" $F_{_{\!S}}$ depend on the odd generators $v_{\pm}$.
Substituting (\ref{lume11}) into (\ref{lume9}) provides the following
twisted cocycle condition for $F_{_{\!S}}$
\begin{equation}\label{lume12}
    F^{12}_{_{S}}(\Delta_{_{\!J}}^{}\otimes1)(F_{_{\!S}}^{})=
F^{23}_{_{S}}(1\otimes\Delta_{_{\!J}}^{})(F_{_{\!S}}^{})\, ,
\end{equation}
where
\begin{equation}\label{lume13}
    \Delta
_{_{\!J}}(a)=F^{}_{_{\!J}}\  \Delta^{{(0)}}(a)\, F_{_{\!J}}^{-1}
\,,
\end{equation}
and  $\Delta^{(0)} (a) = a \otimes 1 + 1 \otimes a$ for  $a\in osp(1|2)$.
Taking into consideration that the twist $F_{_{\!SJ}}$ for small values of $\xi$
should have a form describing classical $r$-matrix (\ref{lume7b})
\begin{equation}\label{lume14}
    F_{_{\!SJ}}=1+\xi(h\otimes e_{+}^{}-v_{+}^{}\otimes v_{+}^{})+{\mathcal O}(\xi^2) \, ,
\end{equation}
one can write the solution of (\ref{lume12}) in the following explicite form:
\begin{eqnarray}\label{lume15}
F_{_{\!S}} &= &1-4\xi\,\frac {v_{+}^{}}{e^{\sigma}+1}\otimes
\frac{v_{+}^{}}{e^{\sigma}+1}
\nonumber \\
&=&
1-\xi\,\frac{v_{+}e^{-\frac{1}{2}\sigma}}
{\cosh\,\frac{1}{2}\sigma}\otimes \frac{v_+ \, e^{-\frac{1}{2}\sigma}}
{\cosh\,\frac{1}{2}\sigma}\ ,
\end{eqnarray}
where $\sigma = \frac{\xi}{2}E_+ = \frac{1}{2}\ln (1 + \xi \, e_+)$
and $\Delta_{_{J}}(\sigma) = \sigma\otimes 1 + 1 \otimes \sigma$. One can show
that
\begin{equation}\label{lume16}
    F_{_{\!S}}^{-1} =
\frac{\cosh\frac{1}{2}\sigma\otimes\cosh\frac{1}{2}\sigma
+\xi\,v_{+}\,e^{-\frac{1}{2}\sigma}\otimes
v_+\,e^{-\frac{1}{2}\sigma}}
{\cosh\frac{1}{2}\Delta_{_{\!J}}(\sigma)}\, .
\end{equation}
Modifying (\ref{lume15}) by a factor $\Phi = \Phi(\sigma)$ as follows
\begin{equation}\label{lume17}
    \widetilde{F}_{_{\!S}}=\Phi\,F_{_{\!S}}\,,
\end{equation}
where
\begin{equation}\label{lume18}
    \Phi=\sqrt{\frac{(e^\sigma+1)\otimes(e^\sigma+1)}
{2(e^\sigma\!\otimes e^\sigma+1)}}\, ,
\end{equation}
one obtains the unitary twist factor, i.e. $\widetilde{F}_{_{\!S}}\,
\widetilde{F}_{_{\!S}}^{*}=
\widetilde{F}_{_{\!S}}^{*}\,
\widetilde{F}_{_{\!S}}=1,
$ provided that the parameter $\xi$ is purely imaginary.
The choice (\ref{lume17}) provides the following deformed coproducts
of $U_\xi (osp(2|1))$

\renewcommand{\theequation}{19\alph{equation}}
\setcounter{equation}{0}
\begin{eqnarray}
\label{lume19a}
{\widetilde{\Delta}}_{_{SJ}}(h)\!\!&=&\!\!h\otimes e^{-2\sigma}\!\!
+1\otimes h+
\xi v_{+}^{}e^{-\sigma}\!\otimes v_{+}^{}e^{-2\sigma}\,,
\\[9pt]
\label{lume19b}
{\widetilde{\Delta}}_{_{SJ}}(v_{+}^{})\!\!&=\!\!&v_{+}^{}\otimes1+
e^{\sigma}\!\otimes v_{+}^{}\,,
\\[9pt]
\label{lume19c}
{\widetilde\Delta}_{_{SJ}}(v_{-}^{})
&= & v_{-}^{}\otimes e^{-\sigma}\!+
1\otimes v_{-}^{}\!+\displaystyle{\frac{\xi}{4}}\biggl\{\!\Bigl(
\bigl\{h,e^{\sigma}\bigr\}\otimes v_{+}^{}e^{-2\sigma}
\nonumber\\[9pt]
&&\ - \{h,v_{+}^{}\}\otimes(e^{\sigma}-1)e^{-2\sigma}
\nonumber \\[12pt]
&& + \ 2\,v_+{}^{}\!\otimes h-\Bigl\{h,\displaystyle{
\frac{v_{+}e^{\sigma}}{e^{\sigma}\!+1}}\Bigr\}\!\otimes\!(e^{\sigma}\!-1)
e^{-\sigma}\!
\nonumber\\[9pt]
&&
+(e^{\sigma}\!-1)\!\otimes\!\Bigl\{h,\displaystyle{\frac{v_{+}}
{e^{\sigma}\!+1}}\Bigr\}\Bigr),\displaystyle{\frac{1}{e^{\sigma}\!\otimes
e^{\sigma}\!+1}}\biggr\}\, ,
\end{eqnarray}
where
\renewcommand{\theequation}{\arabic{equation}}
\setcounter{equation}{19}
\begin{equation}\label{lume20}
    \Delta_{_{\!SJ}} = \widetilde{F}_{_{\!S}} \, F_{_{\!J}}
    \Delta^{(0)}\, F_{_{\!J}}^{-1} \,  \widetilde{F}_{_{\!S}}^{-1} \, .
\end{equation}
Besides we get

\renewcommand{\theequation}{21\alph{equation}}
\setcounter{equation}{0}
\begin{eqnarray}\label{lume21a}
{\widetilde{S}}_{_{\!SJ}}(h)\!\!&
=\!\!&-h\,e^{2\sigma}+\frac{1}{4}(e^{2\sigma}-1)\,,
\\[3pt]
{\widetilde{S}}_{_{SJ}}(v_{+}^{})\!\!&=&\!\!-e^{-\sigma}v_{+}^{}\,,
\\[1pt]
{\widetilde{S}}_{_{SJ}}(v_{-}^{})\!\!&=&\!\!- v_{-}^{}\,e^{\sigma}+
\xi\,h\,v_{+}^{}e^{\sigma}-\frac{\xi}{4}\,v_{+}^{}e^{\sigma}\, ,
\end{eqnarray}
where the formula (\ref{lume4b}) with $F=\widetilde{F}_{_{S}} \, F_{_{J}}$ has been
used.

Following the general framework of twist quantization the universal
 $R$-matrix is given by the formula $(F^{21} \equiv \sum\limits_{i}
 f^{(2)}_{i} \otimes f^{(1)}_{i})$
\renewcommand{\theequation}{\arabic{equation}}
\setcounter{equation}{21}
\begin{equation}\label{lume22}
R = {\widetilde{F}}_{_{\!S}}^{21}\,
{{F}}_{_{\!J}}^{21}
\,
{{F}}_{_{\!J}}^{-1}
\, {\widetilde{F}}_{_{\!S}}^{-1}\,
={\widetilde{F}}^{\, 21}_{_{\!S}}
R_{_{\!J}}^{}{\widetilde{F}}^{-1}_{_{\!S}}\, ,
\end{equation}
where $R_{J}$ is the Jordanian $R$-matrix describing the quantum algebra
$U_\xi (sl(2))$:
\begin{equation}\label{lume23}
R_{_{\!J}}=F^{21}_{_{\!J}}F^{-1}_{_{\!J}}=e^{2\sigma \otimes h}
e^{-2h\otimes \sigma}~.
\end{equation}
It can be added that

a) We discuss here the complex Lie superalgebra $osp(1|2)$. It appears that
one can consider also the Jordanian deformation
 $U_\xi(osp(1|2;R)$ of the real form of $osp(1|2)$, in a way consistent with the
real form of $sl(2)$ providing $sl(2;R) \simeq o(2,1)$ \cite{bor12}.

b) The real form of $U_\xi (osp(1|2))$ extending supersymetrically the algebra
 $U_\xi(o(2,1))$ describes deformed D=1 conformal superalgebra
   \cite{bor15,bor16}.
  It appears
 that possibly for the physical applications it is useful to use the new
 basis in $U(osp(1|2;R))$, with deformed  $osp(1|2;R)$ superalgebra relations (see e.g.
 \cite{bor21,bor22}).

 \section{Beyond $osp(1|2)$}

 The next case of Jordanian deformation which is of physical interest in the
 $osp(1|2n)$ serie is n=2 \cite{bor17},
  providing new
 quantum deformation of graded anti-de-Sitter algebra
 \cite{bor9}.
  In such a case using the Cartan-Weyl basis of $osp(1|4)$ (see also \cite{bor8},
 where the notation is explained)

\begin{eqnarray}\label{lume24}
&& (a)\; the\; rising\; generators:\;\;
e_{1-2}^{},\,\,\,e_{12}^{},\,\,\,e_{11}^{},\,\,\,e_{22}^{},
\,\,\,e_{01}^{},\,\,\,e_{02}^{}~;
\nonumber \\[3pt] 
&& (b)\; the\; lowering\; generators:\;\;
e_{2-1}^{},\,\,\,e_{-2-1}^{},\,\,\,e_{-1-1}^{},\,\,\,e_{-2-2}^{},
\,\,\,e_{-10}^{},\,\,\,e_{-20}^{}~;
\nonumber \\[3pt] 
&& (c)\; the\; Cartan\; generators:\;\;
h_{1}^{}:= e_{1-1}^{},\,\,\,h_{2}^{}:= e_{2-2}^{}~ ,
\end{eqnarray}
 one can write the following general $r$-matrix with its support in Borel
 sub-sueralgebra
\begin{eqnarray}
&& r^{}(\xi_1^{},\xi_{2}^{})\, = \,r_{1}^{}(\xi_1^{})+
r_{2}^{}(\xi_2^{})~, \label{lume25}
\\
&& r_{1}^{}(\xi_1^{}) = \xi_1^{}\Bigl(\frac{1}{2}\,e_{1-1}^{}\wedge
e_{11}^{}+e_{1-2}\wedge e_{12}-2e_{01}^{}\otimes e_{01}^{}\Bigr)~,
\label{lume26}
\\[5pt]
&&
r_{2}^{}(\xi_2^{})= \xi_2^{}\Bigl(\frac{1}{2}\,e_{2-2}^{}\wedge
e_{22}^{}-2e_{02}^{}\otimes e_{02}^{}\Bigr)~. \label{lume27}
\end{eqnarray}
 The twist quantization generated by the classical $r$-matrix (\ref{lume25}) has
 the form \cite{bor8,bor17}

 \begin{equation}\label{lume28}
    F(\xi_{\, 1} , \xi_{\, 2} ) = {\widetilde{F}}_{\, 2}(\xi_{\, 2}) \,
     F_{\, 1} (\xi_{\, 1}) \, ,
\end{equation}
where $F_1$ is the twisting two-tensor corresponding to the classical $r$-matrix
(\ref{lume26}) and $\widetilde{F}_2$ is the two-twisting tensor corresponding to
the $r$-matrix (\ref{lume27}) with generators modified by suitable similarity
map $\widetilde{e}_{ik} = \omega_{\xi_1} \, e_{ik} \, \omega^{-1}_{\xi_1}$
where
\begin{equation}\label{lume29}
\omega_{\xi_1} = \exp\left(
\frac{\xi\, \sigma_{11} \, e_{1-2} \, e_{12}}{1 - e^{2\sigma_{11}}}
\right)
\, \exp \left(
\frac{1}{4} \, \sigma_{11}
\right)\, ,
\end{equation}
and $\sigma_{11}= \frac{1}{2}\ln(1+ \xi_1 e_{11})$.

The 10 bosonic generators $e_{mn}$ (see (\ref{lume24}) if
$m,n= \pm 1, \pm 2$ describe the AdS $O(3,2)$ generators, and the
generators $e_{0m}$ ($m=\pm 1, \pm 2$) define four odd supercharges. Introducing the
AdS radius $R$ and performing the limit $R\to \infty$ one can show \cite{bor9}
  that the classical
 $r$-matrix (\ref{lume25}) has the finite limit if $\xi = \xi_1 = \xi_2$
  and $\xi$ depends on $R$ in the following way
  \begin{equation}\label{lume30}
    \xi(R) = \frac{i}{\kappa \, R}\, .
\end{equation}
 In particular  one obtains in the limit $R\to \infty$ from the classical $r$-matrix
  $r(\xi(R), \xi(R))$ (see (\ref{lume25}))
the following super-Poincar\'{e} classical $r$-matrix:
\begin{equation}\label{lume31}
    r^{SUSY}_{\kappa} =\frac{1}{\kappa}
    r^{LC} + \frac{2}{\kappa}\left(
    Q_1\,  \wedge \, Q_1 + Q_2 \, \wedge \, Q_2 \right) \, ,
\end{equation}
where $Q_m = \lim\limits_{R\to \infty} (i \, R)^{- \frac{1}{2}}\, e_{0m}$
 $(m=1,2)$ and $r^{LC}$ describes the light-cone $\kappa$-deformation of
Poincar\'{e} algebra \cite{bor18,bor19}.
It apears that such a contraction limit
 $R\to \infty$ can be applied also to the twisted coproducts and twisted
  antipode of $U(osp(1|4))$ what provides new deformation of $D=4$
  Poincar\'{e} superalgebra.

In conclusion we would like to state that one
  can introduce by the
contraction procedure two $\kappa$-deformations of $D=4$, $N=1$ supersymmetries

\renewcommand{\fboxsep}{0in}
\renewcommand{\fboxrule}{.1pt}
\newcommand{\rcurltag}{}
\unitlength=0.1cm
{
{\begin{center}
\hskip54pt
{\begin{picture}(89.6,8.32)(7.36,117.12)
\put(-25.36,125.12){\makebox(0,.32)[lb]{Drinfeld-Jimbo}}
\put(25.84,123.44){\vector(1,0){38.4}}
\put(68.4,125.12){\makebox(0,0)[lb]{Standard $\kappa$-deformed $D=4$ }}
\put(-25.36,118.64){\makebox(0,0)[lb]{deformation
$U_q(\mathfrak{osp}(1,4))$}}
\put(29.08,115.96){\makebox(0,0)[lb]{$(q = \frac{1}{\kappa R}; \quad R\to \infty ) $}}
\put(68.4,118.64){\makebox(0,0)[lb]{Poincar\'{e} superalgebra [22]}}
\end{picture}
\rcurltag }
\\[30pt]
\renewcommand{\fboxsep}{0in}
\renewcommand{\fboxrule}{.1pt}
\unitlength=0.1cm
\hskip54pt
{\begin{picture}(89.6,8.32)(7.36,117.12)
\put(-25.36,124.12){\makebox(0,.32)[lb]{Jordanian type}}
\put(25.84,123.44){\vector(1,0){38.4}}
\put(68.4,124.12){\makebox(0,0)[lb]{Light-cone $\kappa$-deformation}}
\put(-25.36,116.64){\makebox(0,0)[lb]{deformation
$U_{\xi_1,\xi_2}(\mathfrak{osp}(1|4))$}}
\put(26.08,115.96){\makebox(0,0)[lb]{$(\xi_1=\xi_2 = \frac{i}{\kappa R};
\ R \to \infty)$}}
\put(74.4,118.64){\makebox(0,0)[lb]{of $D~=~4$ Poincar\'{e} 
}}
\put(74.4,112.64){\makebox(0,0)[lb]{superalgebra}}
\end{picture}
\rcurltag }
\end{center}
}

More detailed description of the second deformation is  provided
by the authors of the present report in \cite{bor9}.

\section*{Acknowledgments}
One of the authors (J.L.) would like to thank
 George S. Pogosyan and Bernardo K. Wolf for his
 warm hospitality at Cocoyoc. The third author (V.N.T.) acknowledges the support from
  the grants RFBR-02-01-00668 and INTAS OPEN 03-01-00837.

\end{document}